# Dynamical Simulation of On-axis Transmission Kikuchi and Spot Diffraction Patterns, Based on Accurate Diffraction Geometry Calibration

Tianbi Zhang (0000-0002-0035-9289) [1], Raynald Gauvin (0000-0003-2513-3128) [2], Aimo Winkelmann (0000-0002-6534-693X) [3,4], T. Ben Britton (0000-0001-5343-9365) [1]

[1] Department of Materials Engineering, The University of British Columbia, 309-6350 Stores Road, Vancouver BC, V6T 1Z4, Canada

[2] Department of Mining and Materials Engineering, McGill University, 3610 University St, Montreal, Quebec, H3A 0C5, Canada

[3] AGH University of Krakow, Academic Centre for Materials and Nanotechnology (ACMiN), al. A. Mickiewicza 30, 30-059 Kraków, Poland.

[4] ST Development GmbH, Wilhelmshöhe 7, 33102 Paderborn, Germany

## Highlights

- Diffraction geometry calibration routine of an on-axis TKD geometry
- Geometric simulation of both Kikuchi bands and diffraction spots
- Full contrast dynamical simulation of on-axis TKD patterns with many diffraction features replicated
- Unlocks enriched 4D-STEM-in-SEM microstructure analysis

**Abstract**

Transmission Kikuchi diffraction in the scanning electron microscope has gained popularity as a materials characterization technique for its high throughput and nanometer-level spatial resolution. While conventional diffraction pattern analysis routines focus on Kikuchi bands on the diffraction patterns, the full physical picture of electron scattering and diffraction pattern formation is more complex. Analysis that accounts for additional diffraction features such as diffraction spots and excess-deficiency effects should provide more robust and accurate indexing, if they can be incorporated in pattern indexing or simulation routines. A more accurate understanding of their physics of formation and geometry is required to enable this change. In this work, we demonstrate geometric and full contrast dynamical simulation of on-axis transmission Kikuchi patterns, based on experimental patterns captured using a modular, direct electron detector-based set-up in the scanning electron microscope. First, a diffraction geometry calibration routine is proposed based on the electron channeling pattern of the direct electron detector. This allows us to accurately account for the position of diffraction spots in both geometric and dynamical simulations with good agreement with experimental patterns. Further, by introducing appropriate weight factors, simulation of incoherent diffuse intensity, and calculation of the energy spectra of diffracted electrons, simulated patterns can be obtained which accurately capture the many diffraction features on experimental patterns. Workflows and findings of this work can be used to improve pattern indexing routines, as well as the understanding of the physical processes in the formation of on-axis transmission Kikuchi patterns.

## Introduction

Kikuchi diffraction techniques such as electron backscattered diffraction (EBSD) and transmission Kikuchi diffraction (TKD) are popular orientation microscopy techniques to study microstructures of crystalline materials [1,2]. Specifically, TKD-in-SEM on electron transparent samples, including off-axis, on-axis and near-axis variants, can offer superior spatial resolution in the order of a few nanometers [3–7], and it has also gained popularity thanks to the development of scanning transmission electron microscopy (STEM)-in-SEM techniques. Furthermore, it is worth noting that the on-axis TKD experiment is very similar in practice to so-called '4D-STEM' (in SEM) [8].

The core piece of Kikuchi diffraction techniques including TKD is capture, processing and analysis of Kikuchi patterns to obtain structural and orientation information of the sample which is encoded in the Kikuchi patterns. To develop accurate and robust pattern analysis routines, a reasonable understanding of the pattern formation geometry and physics is desired, and additional experimental aspects such as pattern centre (PC) calibration, gnomonic distortion, etc. should be considered so as to further improve indexing accuracy.

As a first order approximation, a Kikuchi pattern contains near-straight Kikuchi bands with raised intensities, where each band is a direct space projection of a corresponding lattice plane. Formation of the bands involves a "channeling-in, recoil and channeling-out” process [9–12] of the incident electrons. In the “channeling-in” process, electrons can be seen as being focused to atomic sites in the crystal through coherent scattering. The localized recoil process involves a momentum transfer between the incoming electron and the atomic nuclei, and a minor energy loss (a few eV). For a collection of electrons this results in (virtual) electron sources localized at atomic sites. This process is quasi-elastic and incoherent [12]. Subsequent Bragg diffraction of these electrons by lattice planes (elastic and coherent), i.e. channeling-out, gives rise to a pair of Kossel cones and the Kikuchi band when captured by a detector, based on a gnomonic projection [13,14]. With advancements in computation power and simulation methods such as multi-slice and Bloch Wave approaches [15–17], Kikuchi patterns can be simulated with high fidelity, capturing many intricate features seen in experimental patterns (e.g. higher order bands, complex zone axes and high-order Laue zone rings). These simulation methods have enabled practical dictionary indexing or pattern matching routines for pattern indexing [18,19], and serve as a solid starting point of deep learning algorithms for orientation determination and phase identification [20,21].

The actual physical picture of Kikuchi diffraction is more complicated, because not all incident electrons will undergo the “channeling-in, recoil and channeling-out” process. Those scattering events not localized at atomic sites will generally contribute to a diffuse background, with intensities gradually reducing with increasing scattering angles [22]. Diffraction of this diffuse intensity contributes to excess-deficiency (E/D) lines [23]. Moreover, for on-axis TKD with a suitably thin sample [24], Kikuchi bands can co-exist with diffraction spots which originate directly from coherent scattering from the incident electron beam. More complex dependency of pattern contrast, such as the presence (or absence) of diffraction spots, band contrast inversion and absorption have also been studied and reported [24,25]. We note that practical pattern analysis routines often favour simplicity, and often makes use of only one group of features while neglecting the other, for example, detection and indexing of bands [26], and pattern matching based on simulations with only Kikuchi bands or diffraction spots [18,27].

Nevertheless, these features can be important in interpretation of the diffraction pattern, as they describe the same diffracting crystal in the direct space and reciprocal space respectively, especially when higher angular resolution is desired. For example, Winkelmann et al. showed that by considering E/D effect in dynamical simulation, the accuracy of PC and orientation determination can be improved for EBSD and on-axis TKD patterns (TKPs) [17], and consideration of the diffraction spots should offer similar improvements. Simulation of these features are also useful for understanding the physical processes involved in pattern formation. It is also noticed that while simple kinematic (or geometric) simulations are common for Kikuchi bands and diffraction spots (e.g. [28,29]), it should also be possible to develop models, whether geometric, kinematic or dynamical, which brings the two sets of diffraction features together within the same "framework".

For dynamical simulation aiming to reproduce the full contrast on experimental on-axis TKPs, it is acknowledged that a comprehensive multi-scattering model (e.g. [30,31]) can be overcomplicated and computationally costly. Some simplifications can be applied, for example, by simplifying the treatment of thickness contribution, along with using separate models for simulation of different diffraction features and calculation of energy spectra of scattered electrons, as well as introducing phenomenological weight factors to assemble the full contrast pattern. For example, Winkelmann et al. introduced effective weight factors for the diffuse background pattern to simulate Kikuchi patterns with E/D effect [17]. Shi et al. combined Monte Carlo simulation with dynamical simulation (bands only) to generate "polychromatic" EBSD patterns [32]. Both methods were able to improve pattern matching-based indexing. It is thus anticipated that these methods can be extended and combined to simulate the more complex on-axis TKPs.

We also note that a reasonable simulation also requires accurate knowledge of the diffraction geometry. It is well known that uncertainties in the pattern centre (PC) of Kikuchi patterns can strongly affect the accuracy of orientation determination and high angular resolution methods [33,34]. When diffraction spots are considered as well, the detector tilt will play an additional role as the centre of the diffraction spots (which is the incidence of the direct transmitted beam on the detector) and the PC no longer coincide. This creates an additional challenge to indexing and pattern simulation routines, but in turn proper treatment of detector tilt would then improve the accuracy of pattern analysis, since detector tilt can be ever-present in practice.

In this work, we first present a new calibration routine of an on-axis TKD geometry using electron channeling patterns (ECPs) of a direct electron detector (i.e. the detector is the sample for capturing the ECP), and experimental TKPs for accurate determination of the detector tilt. This serves as a rigorous basis for pattern interpretation and simulation. Then a geometric simulation method is developed by calculating Kikuchi bands and diffraction spots within the same framework. Dynamical simulations are then performed first separately for different features (Kikuchi bands, diffraction spots, E/D, beam broadening and the diffuse background) using a single Bloch wave approach-based simulation tool, and then full contrast patterns are obtained using both a phenomenological model and a Monte Carlo simulation-informed weighting method. This work represents one of the first attempts to treat the two sets of diffraction features (Kikuchi bands and diffraction spots) for low kV (i.e. suitable for conditions between ~ 5 kV and 100 kV) within a single framework, which includes geometry calibration as well as kinematic and dynamical simulations of these key diffraction pattern features. Both experimental and simulation methods and outcomes of this work will deepen the community understanding of the formation process of on-axis TKPs and hopefully inspire the development of more advanced indexing and simulation methods.

## Methodology

**Sample and Electron Microscopy.** A sample of thin flakes of $MoO_3$ on a 3 mm Cu TEM grid was purchased from Ted Pella Inc. (California, USA; product number 625). Electron microscopy and on-axis TKD is performed in a TESCAN (Brno, Czech Republic) Amber-X plasma focused ion beam (pFIB) scanning electron microscope (SEM) at 25-30 keV primary beam energy and 1 nA beam current.

A modular stage [35] is used to perform on-axis TKD (Figure 1). It consists of a base stage, an X-Y-Z linear piezoelectric positioning stage (SmarAct GmbH, Germany) and a MiniPIX Timepix3 direct electron detector (Advacam s.r.o., Czech Republic). The detector has a 100 µm thick Si(001) sensor and works in electron counting mode and frame readout for pattern capture. When the pixel array is aligned with the X-Y beam scan directions of the SEM, and in the absence of detector tilt, orientation of the Si sensor can be described by a Euler angle triplet of [0, 0, 45]° (Bunge convention). The full 256 x 256 pixel array (pixel pitch is 55 μm) is cropped to 252 x 252 pixel patterns in post-processing to avoid edge and corner effects. The entire TKD stage is mounted onto the sample stage of the SEM using a pin mount.

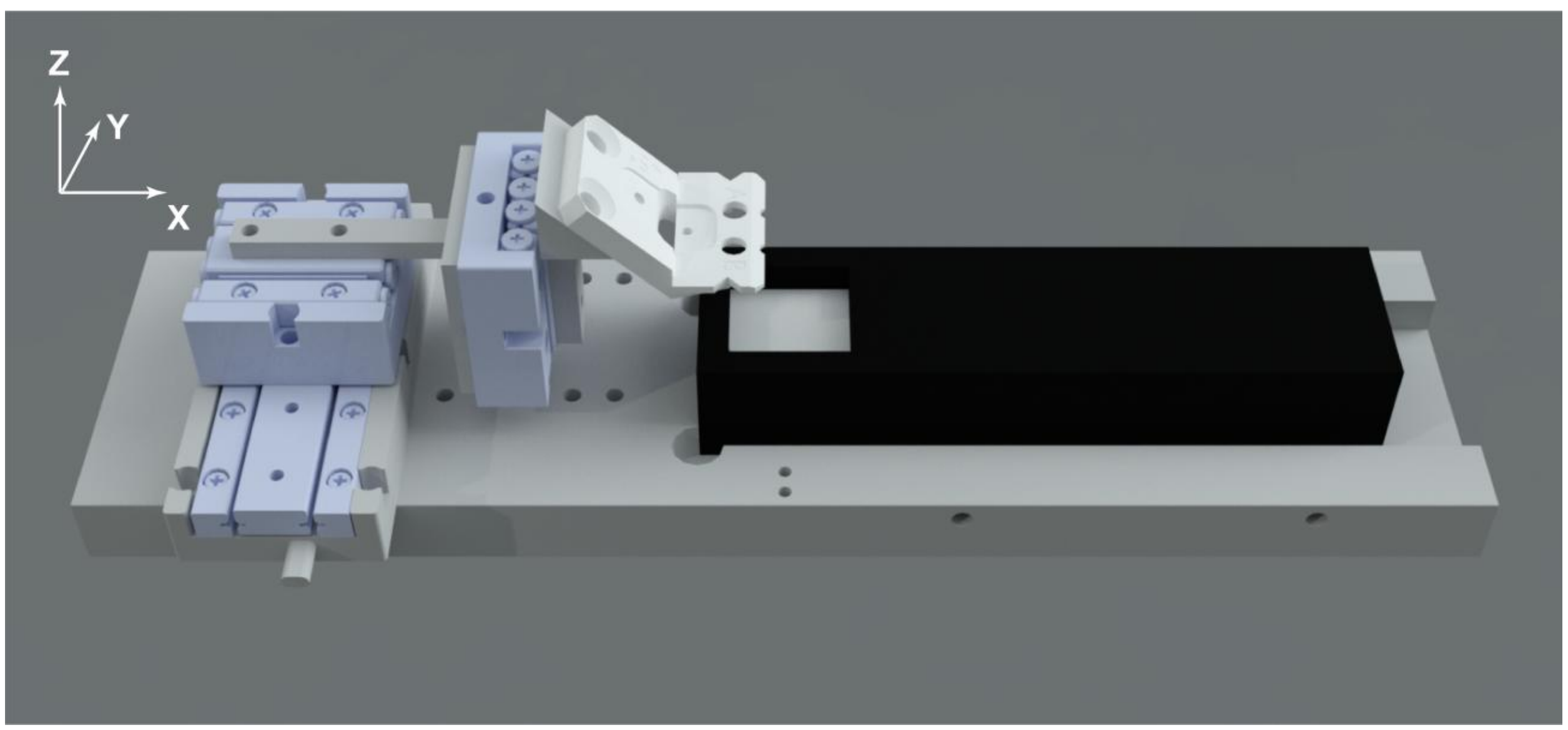

Figure 1. CAD rendering of the on-axis TKD stage.

To capture a TKP, first, the TKD stage is aligned so that the X-Y beam scan grid of the SEM aligns with the X-Y pixel array of the detector. The sample is then brought into the field of view using the piezoelectric stages. After standard SEM focusing and alignment procedures, a TKP can be obtained from a suitable $MoO_3$ flake by pointing the electron beam to a suitable region on the sample. To reduce saturation but increase signal to noise, each TKP is obtained as a sum of 50 frames, with each frame having an exposure time of 0.001 s.

**Diffraction Geometry Calibration.** Calibration of the diffraction geometry focuses on the total vertical distance from the pole piece of the microscope to the detector surface, and the detector tilt.

Two routes are developed for geometry calibration. The first is the "TKP route" by capturing TKPs of the $MoO_3$ sample at different detector distances (DDs), by moving the Z piezoelectric stage in increments of approximately 1 mm intervals (exact interval is read from the control hardware). The same area of the sample is re-centred and the microscope is then re-focused on the sample. Moving the sample in this way only changes the DD. As the horizontal field of view (c. 5 μm) is smaller than 1/10 of the physical pixel size of the detector, shift of patterns due to inaccuracies in re-centering the sample should be negligible. After each change of the DD, the sample is allowed to settle for 1 minute before capturing a TKP. The vertical (Z) position of the SEM stage remains the same during TKP capture.

Focusing, capturing an SEM micrograph of the sample and indexing the TKP (including the PC) provide a measure of the total vertical distance The total distance from the beam rocking point to the sensor is expressed as the following:

$$Z_{\mathrm{TKD}} = \mathrm{WD}_{\mathrm{TKD}} + \frac{\mathrm{DD}_{\mathrm{TKD}}}{\cos(\phi)} \tag{1}$$

where $\mathrm{WD}_{\mathrm{TKD}}$ is the working distance (in mm) to the TKD specimen, $\mathrm{DD}_{\mathrm{TKD}}$ is DD of the TKP sample (in mm) and the detector tilt angle $\phi$ is used to convert the perpendicular distance to the vertical distance to the sensor.

The second route is the electron channeling pattern ("ECP") route, uniquely enabled thanks to the single crystal sensor of the DED. This route is necessary to determine both the DD and the detector tilt. To capture ECPs of the DED sensor, the sample is first moved away from the field of view of the SEM using the piezoelectric stages. The entire stage is also brought closer to the pole piece to increase the solid angle of the ECP by moving the SEM sample stage. Then the SEM was set to the "channeling" mode and an ECP is captured by a 4-quadrant backscattered electron detector. Beam current is increased to 10 nA for higher pattern quality. ECPs are captured after all TKP captures. A detailed workflow of obtaining ECPs using this SEM instrument can be found in [36].

Similarly, orientation and DD are obtained from indexing the ECP which is analogous to the (EBSD) Kikuchi pattern. To convert normalized DD of the ECP to a physical distance (the working distance, WD, to the detector), a linear relationship obtained by Qaiser et al. [36] measured on the same SEM instrument is used:

$$\mathrm{DD}_{\mathrm{ECP}} = (0.2097 \pm 0.005)\mathrm{WD}_{ECP} + (2.3946 \pm 0.058) \quad (2)$$

The total vertical distance, considering stage movement, is expressed by:

$$Z_{\mathrm{ECP}} = \mathrm{WD}_{\mathrm{ECP}} + \Delta Z_{\mathrm{Vertical}} + \Delta Z_{\mathrm{Horizontal}} \quad (3)$$

$\Delta Z_{\mathrm{Vertical}}$ is the vertical shift for moving the stage from TKD position to ECP position. This is directly calculated from the stage Z positions which are included as metadata of the SEM micrographs. $\Delta Z_{\mathrm{Horizontal}}$ is the offset due to the horizontal SEM stage movement, and the detector tilt. This is calculated from the stage shifts in X and Y directions:

$$\Delta Z_{\mathrm{Horizontal}} = \sqrt{\Delta X_{\mathrm{stage}}^2 + \Delta Y_{\mathrm{stage}}^2}\,\tan\phi \quad (4)$$

where stage shifts in X and Y directions are obtained from the metadata of SEM micrographs.

The relationships above regarding the total vertical distance are also schematically illustrated in Figure 2.

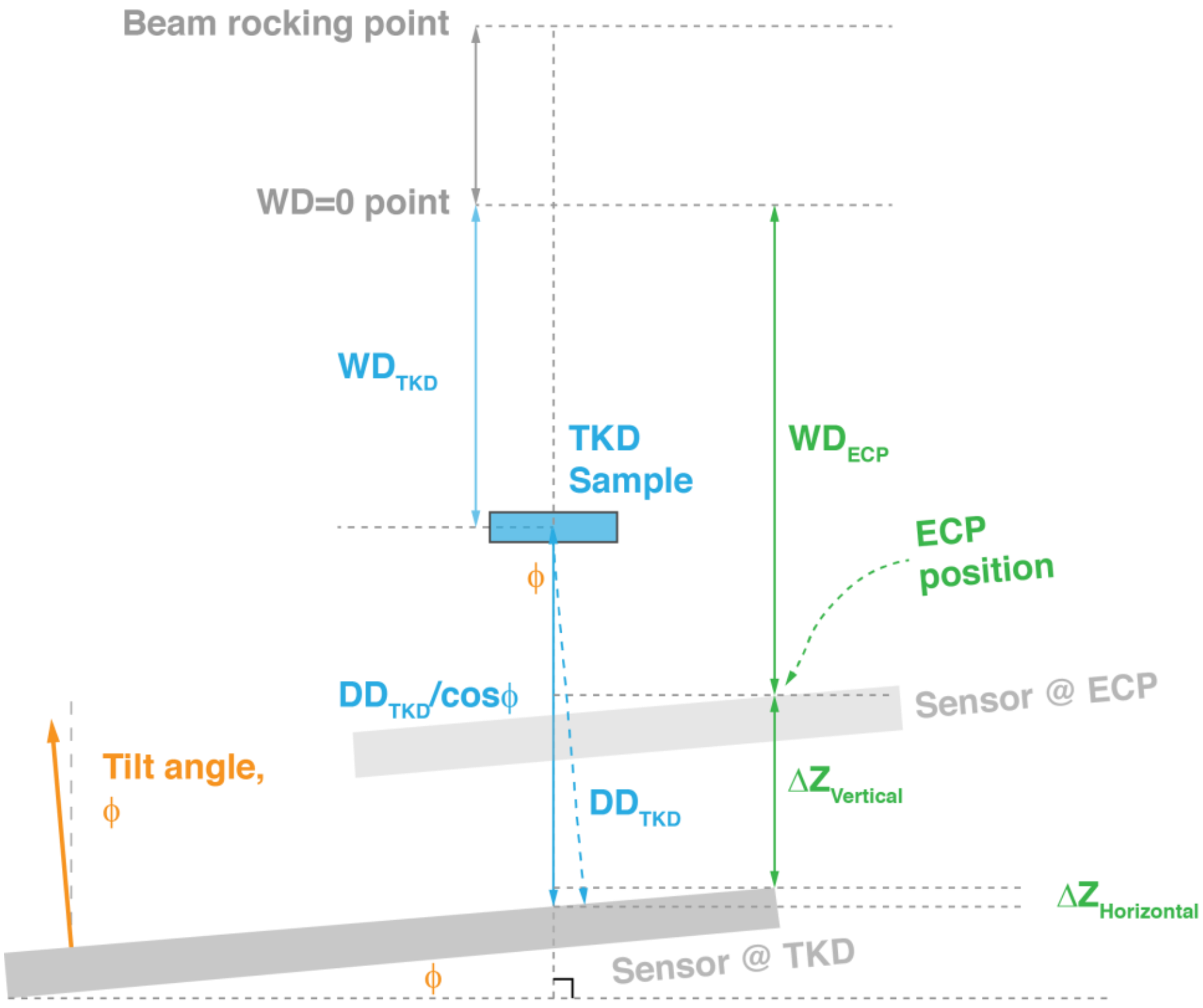

Figure 2. Schematic diagram of the on-axis TKD geometry with a detector tilt for determining the total vertical distance (not to scale). Blue represents the “TKP route” and green represents the “ECP route”.

Tilt of the detector is then calculated from the misorientation between the tilted detector orientation and the reference position. Considering all symmetric equivalents, the rotation with the smallest rotation angle, and the corresponding rotation axis, are taken as the tilt angle $\phi$ and the tilt axis.

**Geometric Pattern Simulation.** Geometric pattern simulation was performed to plot Kikuchi band edges (and/or centre) and diffraction spots in the gnomonic coordinate. The sample ("source point" of the diffraction signal) has a coordinate of (0, 0, 0) and the pattern center (0, 0, 1), and the X and Y coordinates are normalized by the size of the pattern [14]. Miller indices of Kikuchi bands to be plotted are supplied. All symmetry equivalent bands are considered. Kikuchi band centres are simulated based on the geometric relationship that the band centre (i.e. direct projection of the plane) is normal to the corresponding plane normal $\hat{n}_{hkl}$ (a unit vector), so a vector $\vec{d} = (x, y, 1)$ from the sample to a point on the band centre satisfies:

$$\hat{n}_{hkl} \cdot \vec{d} = 0 \tag{5}$$

Similarly, a vector $\vec{d}$ from the sample to a point on the band edge satisfies:

$$\hat{n}_{hkl} \cdot \vec{d} = \pm \sin \theta_{hkl} \tag{6}$$

where $\theta_{hkl}$ is the Bragg angle of the reflection $(hkl)$. Band centre and/or edges can then be plotted using parametric equations. The 3D vector $\hat{n}_{hkl}$ in the gnomonic coordinates is calculated based on the orientation matrix and the crystal structure, and is included in the AstroEBSD toolbox [37].

For simulating diffraction spots, we note that the diffraction spots are centred around the direct transmitted beam (i.e. the (000) point). From Bragg's law, radial position of each diffraction spot $(\mathrm{hkl})$ is $\tan(2\theta_{hkl})$, and the exact position of the spot in the gnomonic coordinate can then be calculated using a parametric form based on the corresponding normal vector (see also Figure 3(a)):

$$X_{hkl} = X_{000} + \tan 2\,\theta_{hkl}\ \ \hat{n}_{hkl,X} \tag{7a}$$

$$Y_{hkl} = Y_{000} + \tan 2\,\theta_{hkl}\ \ \hat{n}_{hkl,Y} \tag{7b}$$

To plot a collection of diffraction spots, the approximate zone axis perpendicular to the pattern plane, and two in-plane crystallographic directions (not co-linear) are calculated based on orientation. These two directions are rounded to nearest integers and used as “bases” to generate a list of reflections by linearly combine the two bases (the maximum number of linear coefficients is a user input). Forbidden reflections, if any, are then excluded from the list, and then the spots are plotted onto the experimental pattern.

Treatment of the detector tilt in this model focuses on calculating the shift of the central point from the PC of the Kikuchi pattern due to the detector tilt, since both the Kikuchi and the spot patterns describe the same orientation. The shift of the direct transmitted beam due to the detector tilt can be calculated using the projection of the rotated unit vector along z (denoted $\vec{z}$) on to the un-tilted horizontal plane (Figure 3):

$$\vec{l} = g_{\mathrm{ECP}}\vec{z} - \frac{\vec{z}\cdot(g_{\mathrm{ECP}}\vec{z})}{(\vec{z}\cdot\vec{z})\vec{z}} \tag{8}$$

where $g_{\mathrm{ECP}}$ is the rotation matrix corresponding to the detector tilt (determined from the ECP route).

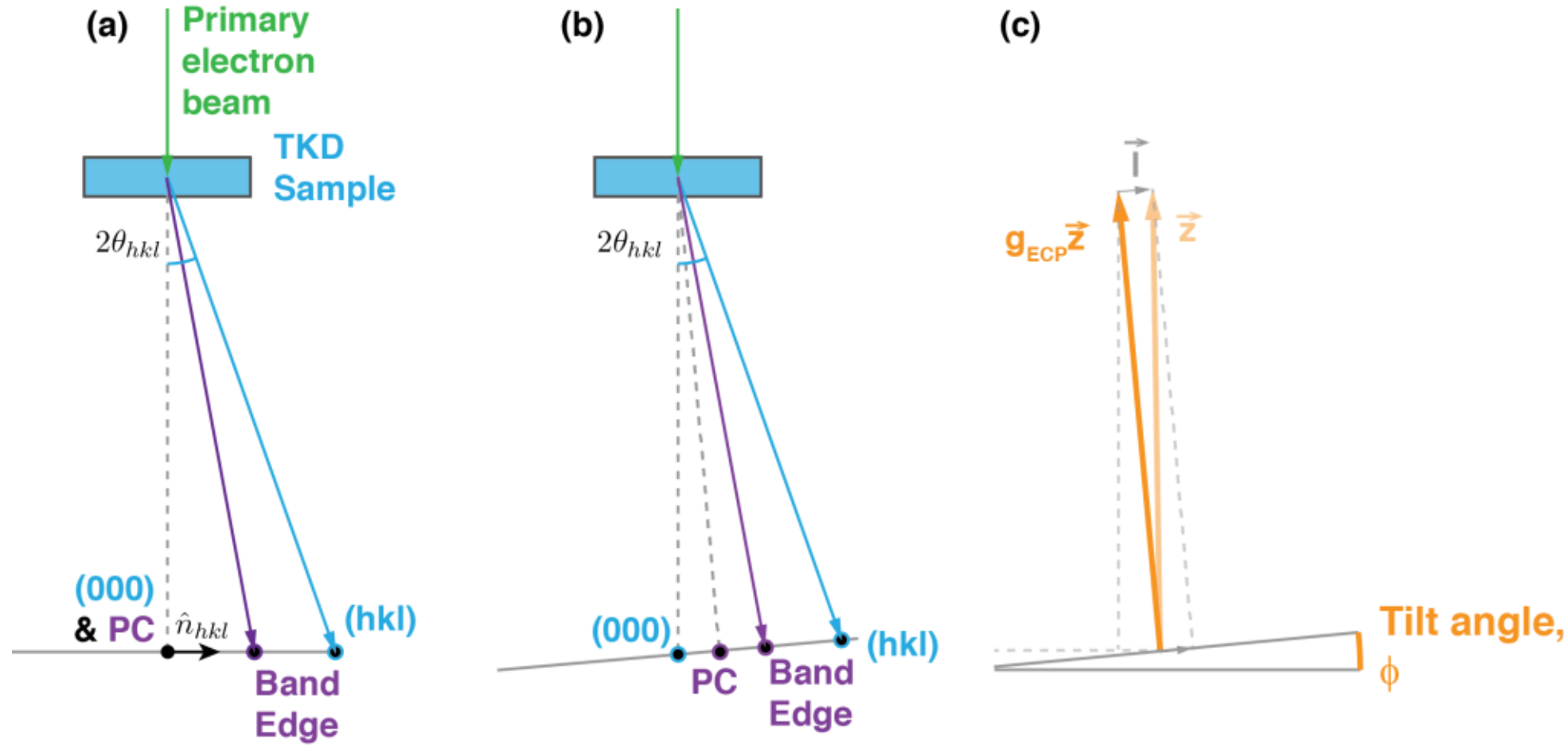

Figure 3. Schematic illustration of a Kikuchi band, a diffraction spot, the pattern centre (PC) of Kikuchi pattern and the centre of the spot diffraction pattern (000): (a) An idealized no-tilt situation; (b) with a detector tilt; (c) the shift vector $\vec{l}$.

Note that the shift from the PC to the new direct transmitted beam position is in the opposite direction, so the new direct beam position is given by:

$$\left(PC_x - l_x, PC_y - l_y, \frac{PC_z}{\cos\phi}\right) \tag{9}$$

Diffraction spots can then be plotted using Equations 7-9, with the radial positions of the spots multiplied by $1/\cos\phi - 1$ due to the small change in DDs.

**Dynamical pattern simulation.** Dynamical pattern simulations are performed using the Bloch wave approach described by Winkelmann et al. [17] implemented in the “BWKD” software package. We simplify the full contrast simulation by using different models within the same package to simulate different diffraction features separately, and then perform a weighted summation of the resulting diffraction patterns:

1. Patterns with Kikuchi band features (with no E/D or diffuse background) are simulated using the coherent point emitter (CPE) sub-model and they are referred to as the “CPE” pattern.
2. Patterns with diffuse background and E/D effects are simulated using the incoherent diffuse intensity (IDI) sub-model and they are referred to as the “IDI” pattern.
3. Patterns with only diffraction spots are also simulated by the IDI sub-model, and they are referred to as the “spot” pattern. Note that this only offers a semi-quantitative simulation with accurate diffraction spot positioning.

PC and orientation of the Kikuchi pattern obtained from indexing are provided to the simulation program. The inelastic mean free path of $MoO_3$ is calculated using the software in reference [38] and extrapolated to 25-30 keV based on a square-root dependency. The sample thickness is assumed to be 100 nm. A summary of the full simulation parameters can be found in supplementary information.

For the IDI model, the detector tilt is incorporated using two phenomenological parameters $u_\varphi$ and $u_\theta$, to describe the direction of the strongest IDI intensity (i.e. the direct transmitted beam) with respect to the perpendicular direction (i.e. PC of the Kikuchi pattern) in a spherical coordinate system. These two parameters are determined from the geometry calibration process and Equation (9).

Simulated CPE and IDI patterns use the same resolution of the experimental patterns (252x252). For simulating diffraction spots, in order to preserve as much as possible the shape and the intensity variation of the diffraction spots and reduce challenges associated with numerical exactness for the simulations, simulations of 1024 x 1024 pixels are first performed. Then these patterns are resized to 252 x 252 using a box-shaped kernel. This approach is similar to the downsampling approach of simulated Kikuchi patterns by Winkelmann et al. [39] for high-resolution EBSD. All patterns have been normalized to have intensities between 0 and 1 (inclusive).

Subsequently, the full pattern contrast simulated pattern is calculated as a weighted sum of different patterns. We present two methods to determine the weight factors, with the detailed approaches explained in the next section.

**Monte Carlo simulation.**

Monte Carlo simulations of the energy distribution of electrons transmitted through and scattered by 100 nm thick $MoO_3$ were performed for scattering angles from 0 to 180 degrees. The continuous slowing down approximation (CSDA) is applied for simplicity. This approximation leads to a minimum energy loss related to sample thickness. Details of the Monte Carlo simulation method can be found in reference [40]. Considering the overall scattering angle and energy distribution, 23 CPE and IDI patterns are simulated between 29.797 and 29.820 keV. Note that these values are the representative of the mean energy of scattered electrons. Subsequently, for calculations using the energy spectrum for each individual pixel and its corresponding scattering angle, the energy spectrum is linearly interpolated.

**Data processing.** Pattern simulations using the "BWKD" package use a Python wrapper. All other data processing routines are performed in MATLAB with the MTEX toolbox (version 5.11.2) [41]. The simulated patterns are saved as text files (as double precision floating point numbers). Indexing of the experimental TKPs is performed using the refined template matching method with PC refinement [19,37]. Indexing of ECPs is performed using the AstroECP tool in AstroEBSD [36]. Dynamical templates of $MoO_3$ and Si are computed by BWKD [17] and EMsoft [42] respectively. All MATLAB scripts are available at https://github.com/ExpMicroMech/AstroEBSD.

## Results

**Diffraction geometry calibration.** ECPs from the 25 and 30 keV TKD experiments, simulated ECPs, as well as a simulated pattern for the reference orientation are shown in Figure 4.

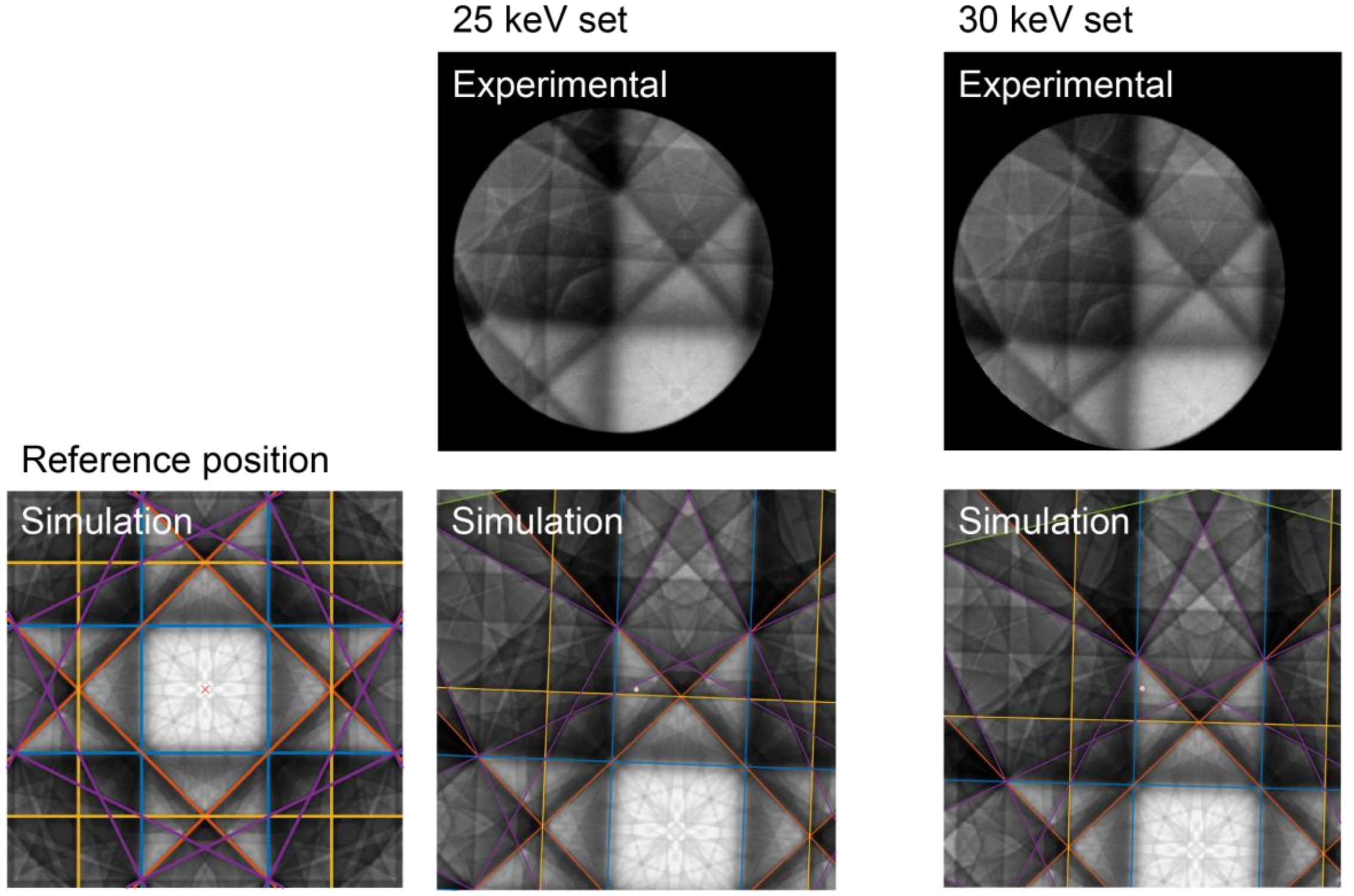

Figure 4. Experimental and simulated electron channeling patterns (ECPs) of the on-axis TKD detector for the 25 keV and 30 keV $MoO_3$ datasets. Note that the ECPs were captured at 20 keV. A simulated ECP corresponding to zero detector tilt are included as reference. Edges of selected bands are also shown.

Figure 5 shows the full and background corrected TKPs captured at 25 and 30 keV primary electron beam energy and the indexed DDs correspond to each pattern.

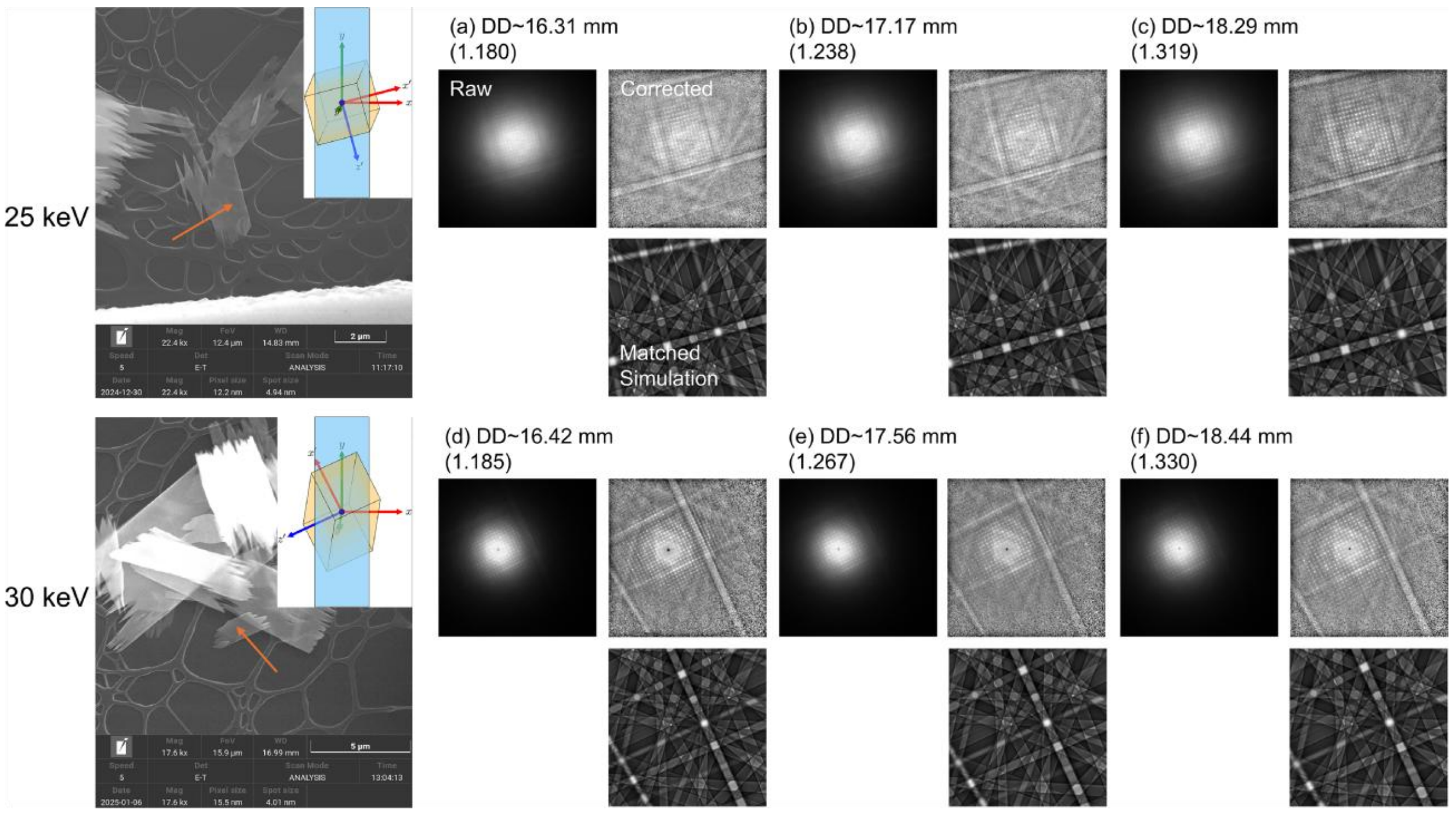

Figure 5. Experimental, corrected and matched on-axis TKPs from $MoO_3$ at 25 keV (a-c) and 30 keV (d-f). SEM micrographs of the $MoO_3$ flakes and crystal shape plots are shown alongside experimental patterns as references. Only Kikuchi bands are simulated for pattern matching and in the matched simulations on this figure.

Table 1 shows the total vertical distances between the beam rocking point and the detector sensor, calculated from the two calibration routes. The TKP and ECP routes show good agreement at 25 and 30 keV cases. Note that the ECP route is associated with larger errors due to the uncertainties in the coefficients of Equation 2.

Table 1. Total vertical distances between the beam rocking point and the detector sensor determined from the TKP and ECP routes of geometry calibration. $Z_{\mathrm{TKD}}$ is reported as 95% confidence interval and errors in $Z_{\mathrm{ECP}}$ are based on Equation 2.

| | Energy (keV) | 25 | | | 30 | | |
|---|---|---|---|---|---|---|---|
| | Tilt angle $\phi$ (°) | 3.53 | | | 3.70 | | |
| | Tilt axis *[uvw]* | [0.761 0.231 -0.607] | | | [0.874 0.302 -0.381] | | |
| TKP Route | TKP Location | 1 | 2 | 3 | 1 | 2 | 3 |
| | $\mathrm{WD}_{\mathrm{TKD}}$ (mm) | 14.83 | 13.82 | 12.79 | 14.47 | 13.46 | 12.42 |
| | $\mathrm{DD}_{\mathrm{TKD}}$ | 1.180 | 1.238 | 1.319 | 1.185 | 1.267 | 1.330 |
| | $\frac{\mathrm{DD}_{\mathrm{TKD}}}{\cos(\phi)}$ (mm) | 16.34 | 17.19 | 18.32 | 16.46 | 17.59 | 18.47 |
| | $Z_{\mathrm{TKD}}$ (mm) | 31.17 | 31.01 | 31.11 | 30.93 | 31.05 | 30.89 |
| | $Average\ Z_{\mathrm{TKD}}$ (mm) | 31.10±0.21 | | | 30.96±0.22 | | |
| ECP Route | $\mathrm{DD}_{\mathrm{ECP}}$ | 7.481 | | | 7.175 | | |
| | $\mathrm{WD}_{\mathrm{ECP}}$ (mm) | 24.26 | | | 22.80 | | |
| | $\Delta Z_{\mathrm{Vertical}}$ (mm) | 6.50 | | | 7.44 | | |
| | $\Delta Z_{\mathrm{Horizontal}}$ (mm) | 0.55 | | | 0.43 | | |
| | $Z_{\mathrm{ECP}}$ (mm) | 31.30±2.12 | | | 30.68±2.03 | | |

**Geometric Simulation.** With detector tilt determined and diffraction geometry calibrated, we can now apply the geometric simulation model to the $MoO_3$ dataset. Examples of 25 keV and 30 keV patterns are shown in Figure 6. Overall, a good agreement can be achieved for both the diffraction spots and the Kikuchi bands. Small mismatches can still be observed, which can be related to uncertainties in orientation and PC determination. The effect of considering the detector tilt on the position of the spot diffraction pattern can be found in Supplementary Information.

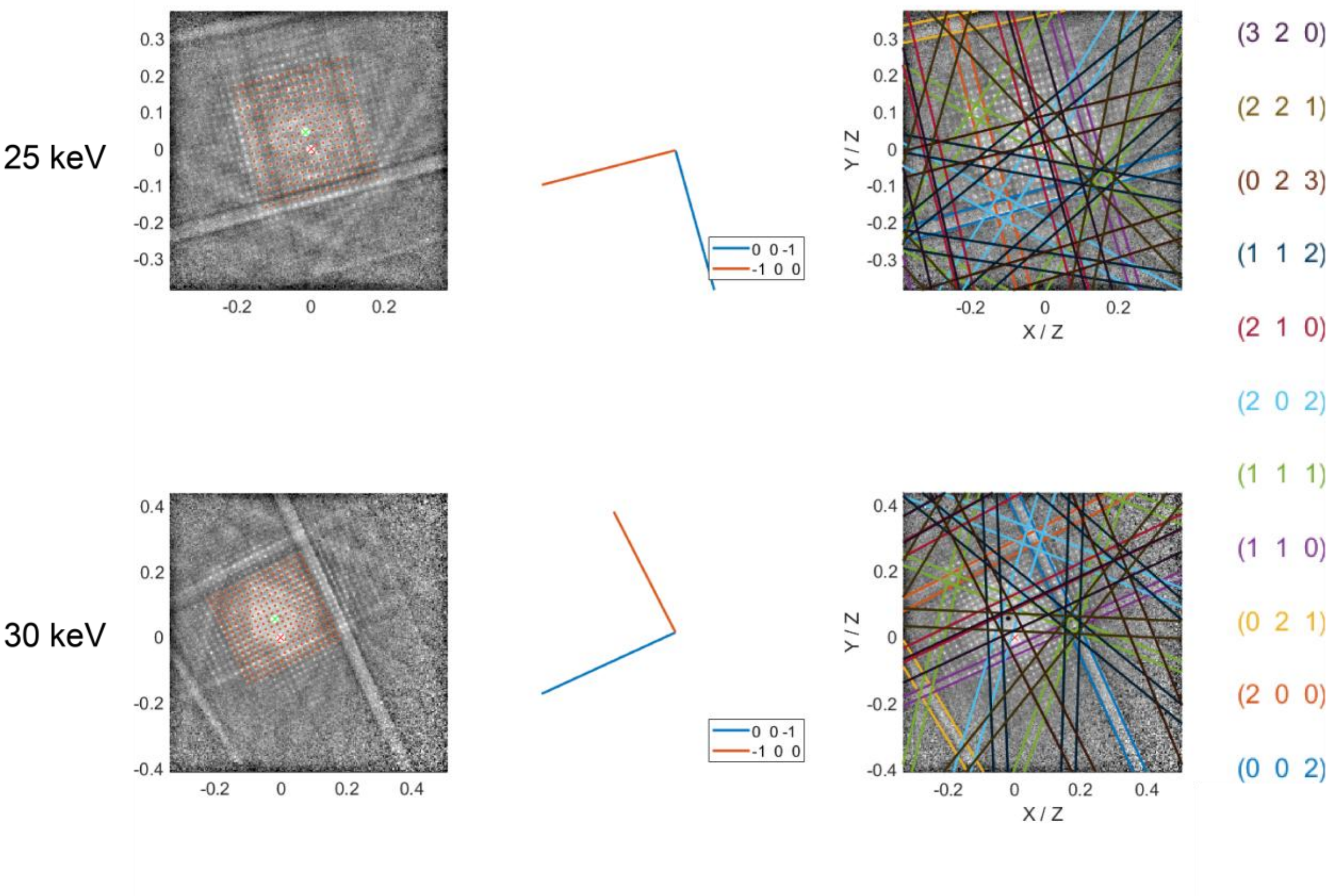

Figure 6. Experimental on-axis TKPs of $MoO_3$ at 25 keV and 30 keV with overlaid band edges and diffraction spots (orange) simulated by the geometric method. Color code of bands is included on the right. (color version online)

**Dynamical Simulation.** The aim of the dynamical simulations is to replicate the full contrast on experimental on-axis TKPs with as many diffraction features considered as possible. Here we use the TKP of $MoO_3$ captured at 30 keV and a DD of 1.185 shown in this manuscript as a benchmark case. The detector tilt translates into $u_{\varphi}$ = 71.64° and $u_{\theta}$ = 2.882°.

Our overall approach, combining different diffraction features (CPE-bands, IDI-diffuse intensity and E/D, and spots), is based on the method by Winkelmann et al. [17] for EBSD patterns, which treats the full contrast pattern as a weighted sum of the CPE and IDI patterns:

$$I_{TKP} = k_{CPE} I_{CPE} + (1 - k_{CPE}) I_{IDI} \tag{10}$$

where $I_{CPE}$ and $I_{IDI}$ are the (intensities of) CPE and IDI patterns respectively, and $k_{CPE}$ is a phenomenological weight factor. For on-axis TKPs, we can expand this formulation to include diffraction spots:

$$I_{TKP} = k_{CPE} I_{CPE} + (1 - k_{CPE}) I_{IDI} + k_{Spots} I_{Spots} \tag{11}$$

Note that the weight of the spot pattern is not coupled with $k_{CPE}$ as the spot simulation is only semi-quantitative.

The IDI model controls the spread of the scattered beam, so it implicitly contains thickness information. The model assumes a Gaussian function for the probability distribution of the outgoing directions with respect to the strongest intensity direction using a phenomenological standard deviation term $\alpha$. A larger $\alpha$ results in a broader, more uniform distribution, and as $\alpha$ decreases, the spread is smaller, and eventually confined to a small range of very similar directions, obtaining diffraction spots. This effect of $\alpha$ on IDI patterns is shown in Figure 7 along with the CPE pattern.

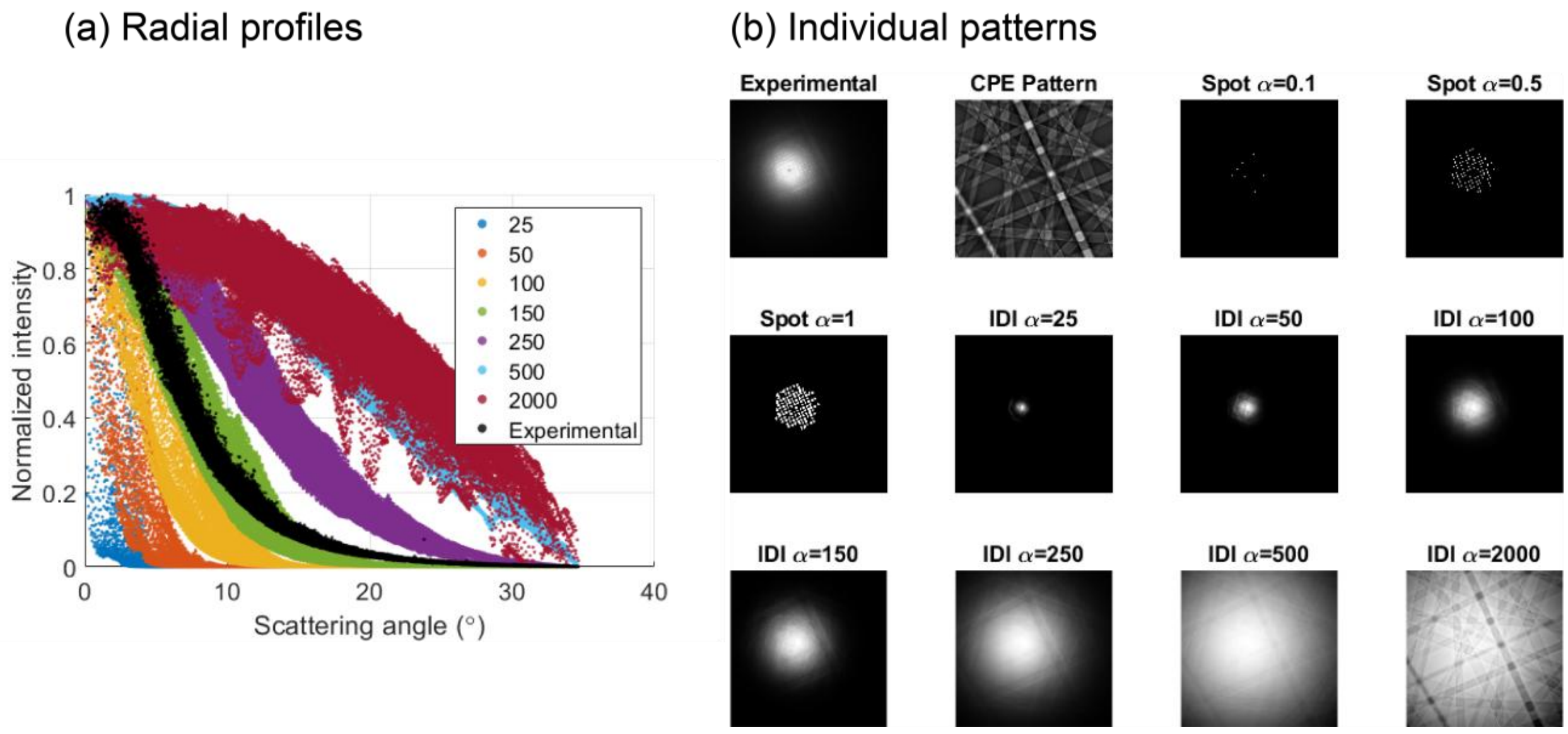

Figure 7. (a) Radial intensity profiles of the experimental pattern and simulated IDI patterns with different $\alpha$. (b) Dynamically simulated patterns showing the effect of $\alpha$ on the IDI pattern.

Based on the radial intensity profiles of the patterns in Figure 7, an $\alpha$ of 150 gives a reasonable match with the experimental pattern. Spot pattern uses $\alpha = 1$.

**Phenomenological approach.** We first present a phenomenological approach by assuming a constant $k_{CPE}$ across the pattern. An optimal value of $k_{CPE}$ can be determined through image-based cross-correlation between experimental and simulated patterns. To reduce the effect of low signal-to-noise ratio at high scattering angles, cross-correlation is performed with a mask applied to all patterns, so that pixels with scattering angles >25° are masked out.

Using the IDI pattern with $\alpha = 150$ and a $k_{Spots} = 0.15$, a very high XCC close to 1 is obtained at $k_{CPE}{\sim}0$, suggesting a minor contribution of the CPE pattern to the full contrast pattern. This qualitatively agrees with the fact that the required “channeling-in, recoil, channeling-out” scattering (or quasi-elastic and incoherent scattering) only accounts for a small portion of the total scattered electrons [22,43]. We choose $k_{CPE} = 0.01$.

For background corrected patterns, while a high XCC can still be observed, the corrected simulated pattern shows a clear transition from the IDI pattern (Figure 8(a)) at lower scattering angles to the CPE pattern at higher scattering angles, which is considered unrealistic. This hints the use of additional IDI pattern(s) to cover high scattering angle regions, and we can adapt a generalized form of Equation 11:

$$I_{TKP} = k_{CPE} I_{CPE} + \sum_i k_{IDI,i}\, I_{IDI,i} + k_{Spots} I_{Spots} \qquad (12)$$

where $k_{CPE} + \sum_i k_{IDI,i} = 1$. Here we present a simple remedy using a second pattern with $\alpha = 500$. This pattern is chosen for appropriate beam broadening which does not cause significant band contrast inversion, in accord with the experimental pattern. We note that this multiple IDI pattern approach qualitatively incorporates thickness effects.

With $k_{CPE} = 0.01$ and $k_{Spots} = 0.15$ from the one IDI pattern result, $k_{IDI,150}$ is determined again from cross-correlation. Based on XCC (Figure 8(b)), $k_{IDI,150} = 0.93$ is chosen. Background corrected patterns in Figure 8(b) does not show the CPE-IDI transition. Meanwhile E/D effects of Kikuchi bands are also accurately captured throughout the pattern, consistent with the experimental patterns. These observations mean that while the low angle scattering term dominates, beam broadening effect at larger scattering angles still accounts for a small but important portion of scattered electrons, especially at larger scattering angles, which results in the observed pattern contrast.

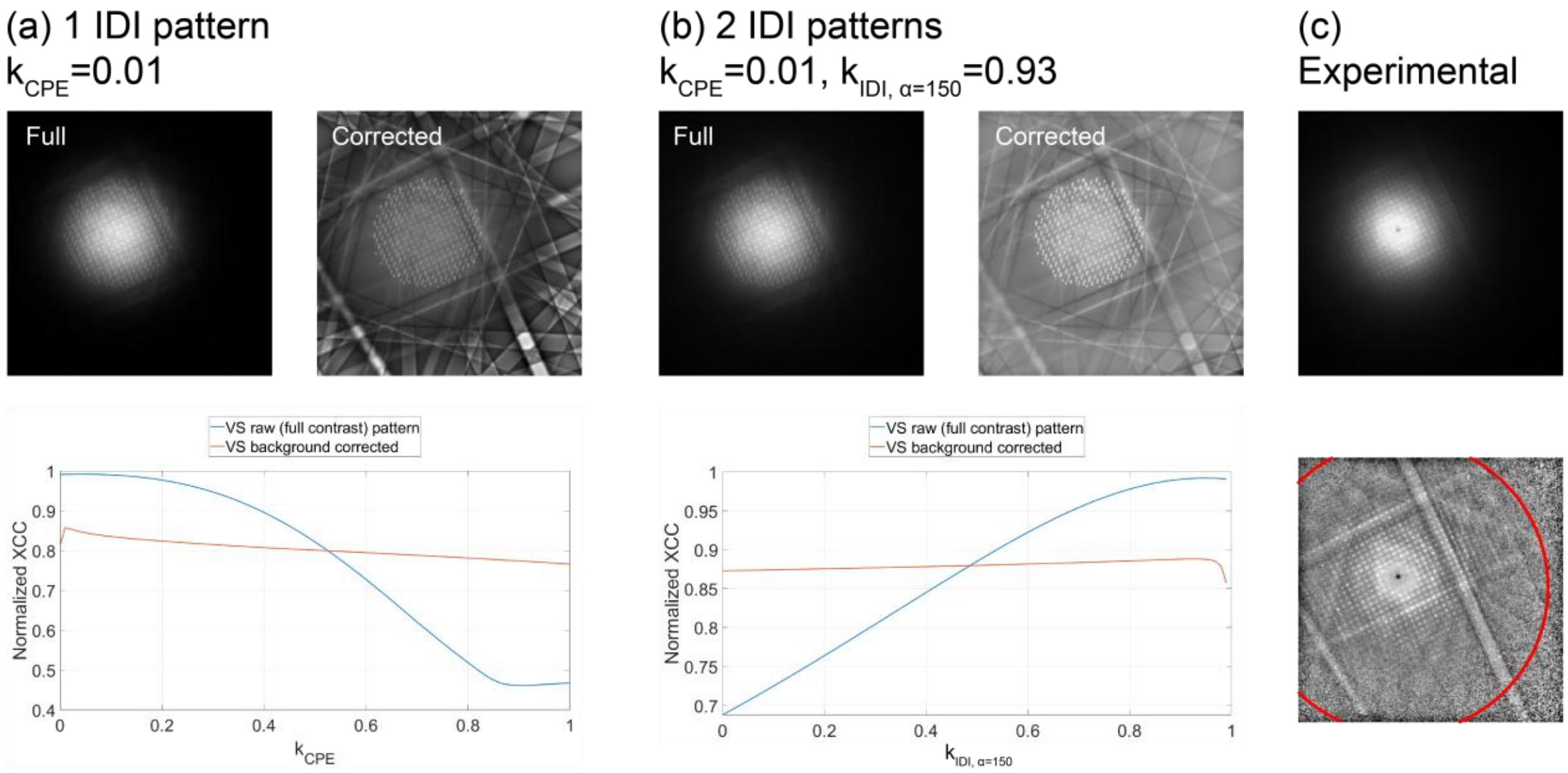

Figure 8. Full contrast, background corrected simulated TKPs and plots of normalized cross-correlation coefficient (XCC) from the phenomenological dynamical simulation approach, using (a) 1 IDI pattern and (b) 2 IDI patterns. Location of the angular mask for XCC is indicated on the experimental pattern (red line).

**Physics-based approach.** The generalized form of the weighted sum (Equation 12) prompts a more physic-based approach. Here we demonstrate a method to create a polychromatic diffraction pattern and determine a position-dependent $k_{CPE}$ from the energy spectra of scattered electrons.

To create the polychromatic CPE and IDI patterns, we follow a method simplified from that shown by Shi et al. for EBSD patterns [32]. The mean energy of scattered electrons at each pixel on the pattern is used to select and fill the corresponding intensity values from the simulated monochromatic pattern. Next, a threshold energy loss is used to separate incoherent (IDI) and coherent (CPE) scattering to determine $k_{CPE}$ at each pixel. The cumulative probability of scattered electrons with energy loss higher than the threshold is $k_{IDI}$, and the rest is considered $k_{CPE}$ so that $k_{CPE} + \sum_i k_{IDI,i} = 1$ is still satisfied. We set this threshold at 0.181 keV. Based on the findings from the phenomenological approach, here we only present results using two IDI patterns ($\alpha = 150, 500$). For simplicity, we further modify Equation 12 such that:

$$I_{TKP} = k_{CPE} I_{CPE} + \sum_i k_{IDI,i} (1 - k_{CPE}) I_{IDI,i} + k_{Spots} I_{Spots} \tag{13}$$

and assume $k_{IDI,i}$ is a constant. XCC analysis showed a similar trend from the phenomenological approach (Figure 9, c.f. Figure 8(b)). Example patterns with $k_{IDI,150} = 0.9$ are shown in Figure 9. The average value of $k_{CPE}$ is 0.079.

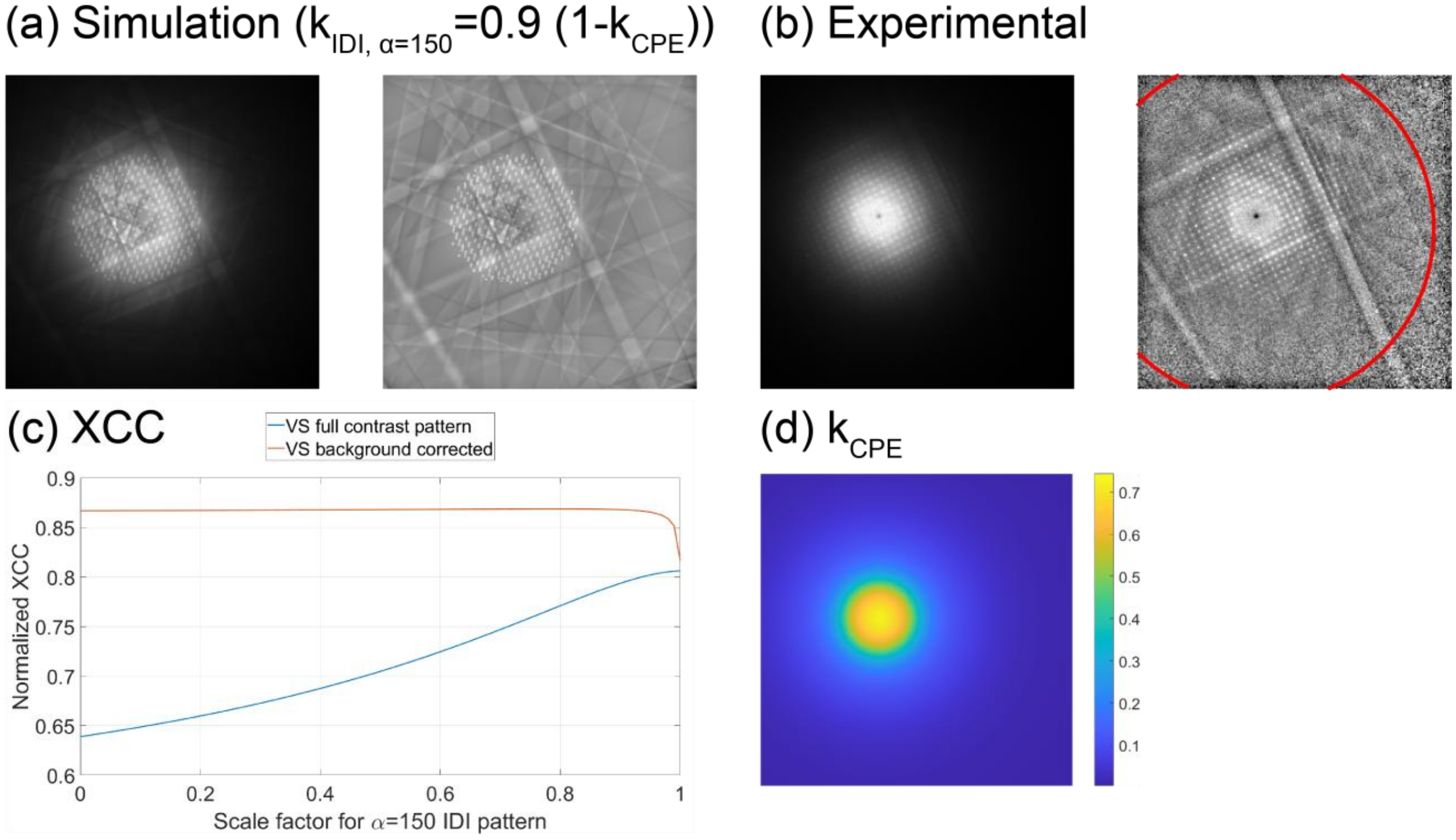

Figure 9. Full contrast and background corrected patterns using the physics-based approach with 2 IDI patterns. Map of $k_{CPE}$ across the pattern are included for reference. Threshold energy loss is 0.181 keV. Location of the angular mask for XCC is indicated on the experimental pattern (red line).

## Discussions

The present work demonstrates the following approaches to improve the experimental workflow and understanding of on-axis TKD: (1) diffraction geometry calibration, (2) geometric pattern simulation, and (3) dynamical simulation to capture full pattern contrast, which may further enhance the capability of TKD (in SEM) methods in materials characterization.

**Geometry calibration.** The present method is developed based on the uniqueness of using a direct electron detector with a single-crystalline sensor, allowing the capture of its ECP and subsequent geometry calibration. We note that compared to the ECPs in works such as Qaiser et al. for electron channeling contrast imaging, ECPs captured in this work are of lower quality due to surface cleanliness of the detector (as the sensor surface cannot be plasma cleaned due to exposed wires) and total pattern collection time. Further, due to the height of the modular TKD stage, the working distance for capturing the ECP is limited to approximately 20-30 mm, and the solid angle captured by the ECP is reduced. This means the indexing accuracy and thus determination of the DD of the ECP can be limited, and the relative error in $WD_{ECP}$ is found to be around 10%. Accuracy of orientation determination from the ECP, and thus the detector tilt, was found to be around 0.1° based on uncertainties of the optical centre of the pattern. [36] Similarly, the accuracy of indexing of the TKPs is also limited by pattern quality, as shorter exposure times were used to reduce sample drift and saturation. Note that the uncertainty of DD for each individual TKP is difficult to evaluate. We also note that, in close connection to the ECP of the detector, geometry calibration may also be performed using the “detector pattern”. These patterns can be captured by the DED from a divergent electron source with no diffraction signals present, and they often appear as a contrast inverted Kikuchi pattern. A geometry calibration for an EBSD geometry using such patterns has been demonstrated by Vespucci et al. [44]

**Geometric simulation method.** Simulation of band edges in addition to the band widths can improve the existing Hough transformation-based indexing methods, the majority of which focus on band centre. While determination of the Bragg angle (band edge) directly from the pattern can be difficult [45], geometric simulation can serve as a first order approximation.

For simulation of the diffraction spots, the main limitation is that the list of diffraction spots calculated will only cover (a subset of) those in the zeroth order Laue zone. For the $MoO_3$ data, due to the larger unit cell, it is expected that all diffraction spots observed and simulated are within the zeroth order Laue zone. The simple model only predicts the central point of each diffraction spot and does not provide any information about the size or shape of each spot.

Our geometry calibration and geometric simulation routine can further improve indexing accuracy, first by reducing the uncertainties associated with the PC [34], and second by making use of the diffraction spots. With the positions of the diffraction spots calculated, we anticipate that a pattern indexing method incorporating both band and spot features can be developed, where band features can be first used as a crude indexing pass and PC determination, and the spot patterns be used for geometry calibration or orientation refinement (or vice versa, using the shift of spot patterns from the PC to back calculate diffraction geometry). For example, a refinement of orientation can be developed by comparing the calculated spot positions with those from peak detection algorithms on the experimental pattern. More accurately locating the diffraction spots can also help with intensity-based analysis (e.g. thickness as in convergent beam electron diffraction methods). Diffraction geometry calibration is then vital for the success of such methods.

**Dynamical simulation.** The main difference between the two full-contrast pattern simulation methods is the treatment of $k_{CPE}$. The phenomenological approach uses a constant $k_{CPE}$ fitted from XCC analysis, while the physics-based method results in a strongly varying $k_{CPE}$ across the pattern (Figure 9) based on per-pixel energy spectrum. Whether it is appropriate to use a constant (or varying) weight factor for each IDI term, or to simply weight the varying $k_{CPE}$ for each IDI pattern, should depend on the nature of the IDI patterns, or thickness-dependent simulated patterns in general. For example, if they represent an infinitesimal thickness element (which favours a constant weight factor as is in an integral), or the sum of contributions from a range of thickness (which should favour a varying weight factor across the pattern). Note that this has been demonstrated in CPE-like simulated patterns [46].

We notice some differences in certain Kikuchi bands between simulation and experimental patterns. First, band contrast of certain bands (e.g. the (002) band) appear higher in the experimental patterns. This is a combined result of the magnitude of $k_{CPE}$ being low, as calculated from the MC results, and the band contrast in the CPE pattern. Second, certain bands which appeared to be “pushed out” from the direct transmitted beam. While Reimer attributed it to a thickness effect [43], many on-axis TKD-in-SEM works, including studies probing thickness effects on pattern contrast, did not [4,24,47]. In the present model, we attribute this effect to using two IDI patterns as the apparent source, which also points to a thickness-based effect. Since the BWKD model is also based on certain simplifications and phenomenological parameters, a more detailed physical explanation of this effect and other thickness-related effects alike should be the subject of future studies.

Some discrepancies are related to the experimental conditions. For example, the amorphous carbon film supporting the $MoO_3$ flakes may contribute to additional beam broadening and blurring of the experimental TKP, as well as noise on the pattern, the uneven signal-to-noise ratio distribution across the pattern, and beam saturation and coincidence loss near the direct transmitted beam (Figure 5), which are not included in the dynamical simulation. In this sense, even with the diffuse background included, dynamical simulation still represents an idealized scenario, so a “perfect match” between experimental and simulated pattern will not be attained. Better experimental patterns without saturation or coincidence loss may be obtained with alternative experimental settings especially detector readout and/or hardware.

A region with slightly reduced intensity near the direct beam can be seen on simulated patterns due to the locally high $k_{CPE}$ values (Figure 9). We note that similar localized contrast inversion can also be seen on experimental patterns as a thickness effect [24,47] and the current approach replicated it in a phenomenological manner.

Although this work only demonstrated a single case study using a single detector geometry, we would like to emphasize that the pattern simulation methodology should be equally applicable to on-axis TKD patterns captured from other crystal structures and using different detector geometries (such as pixel pitch, pixel array size). It is noted that $MoO_3$ is a commonly used calibration standard for (S)TEM and it was chosen for this reason. We anticipate that the effect of thickness and certain detector functionalities, such as energy threshold, may be incorporated by tuning the phenomenological parameters.

**Limitations of the simulation tools.** The spot simulation tool should be the subject of future improvement. First, spot shape effect (see supplementary information for details) forces the approach of simulating a large pattern and down-sampling. In terms of the presence (or absence) and geometry of the spot pattern, currently the simulation tool relies on direct pattern comparison. An improvement of the present model may include a separate model for diffraction spots based on direct pattern comparison, or from empirical studies (based on e.g. material, thickness, beam energy, beam convergence angle) such as the observation from [24], and decide on the appearance and treatment of spot patterns.

We realize the various simplifications used in the MC and dynamical simulations. The use of phenomenological parameters and models in the BWKD simulation means the pattern intensities and weight parameters should be treated semi-quantitatively. We also notice that the effect of incident beam diffraction effect ("channeling-in") is not included. Similarly, the use of the CSDA in the Monte Carlo simulations rather than a discrete model is also a simplification, and we anticipate that a discrete model may make the selection of the threshold energy loss more realistic.

## Conclusion

We presented a series of experimental workflow and simulation methods to improve the capability and understanding of on-axis TKD. First, a diffraction geometry calibration method is established using a direct electron detector for accurate determination of detector tilt angle and axis. Next, a geometric pattern simulation routine for both Kikuchi bands and spots is demonstrated. By making use of the calibrated geometry, the shift of the diffraction spots due to detector tilt can be accurately incorporated, along with the Kikuchi bands.

We also demonstrated, based on the calibrated geometry, methods for full contrast simulation of on-axis TKPs, which implicitly considers thickness effects. By carefully determining the weight factors for patterns containing different diffraction features, the full contrast simulated patterns are able to capture many important features on experimental patterns, including the diffraction spots, E/D effect, and thickness effects such as partial band contrast inversion, with the magnitudes of the weight factors also qualitatively agree with common physical models.  Results from this work can be integrated into indexing method or inspire the development of new indexing and refinement routines, to further improve indexing accuracy, and improve the understanding of the pattern formation process.

**Acknowledgements**

T.Z. and T.B.B. acknowledge the following funding support: Natural Sciences and Engineering Research Council of Canada (NSERC) [Discovery grant: RGPIN-2022-04762, ‘Advances in Data Driven Quantitative Materials Characterisation']; the electron microscopy experiments were performed in the Electron Microscopy Laboratory, in the Department of Materials Engineering, on equipment funded by British Columbia Knowledge Fund (BCKDF) and Canada Foundation for Innovation – Innovation Fund (CFI-IF) [#39798, ‘AM+'] and [#43737, 3D-MARVIN]. T. Z. and T.B.B thank M. Haroon Qaiser (University of British Columbia) for help on capturing and indexing of electron channeling patterns.

### Data availability statement

Experimental patterns will be available on Zenodo upon acceptance of the manuscript. Analysis scripts are available on GitHub at https://github.com/ExpMicroMech/AstroEBSD. The BWKD dynamical simulation engine is available within the AztecCrystal - MapSweeper software (Oxford Instruments), for a Python frontend contact A.W."

.

### CRediT Author Contribution Statement

T. Zhang: data curation, formal analysis, methodology, visualization, validation, writing – initial draft

R. Gauvin: methodology, data curation, formal analysis, methodology, writing – review & editing.

A. Winkelmann: methodology, data curation, formal analysis, methodology, writing – review & editing.

T. B. Britton: conceptualization, methodology, funding acquisition, project administration, resources, supervision, writing – review & editing.

### References

1. Schwartz, A. J., Kumar, M., Adams, B. L., & Field, D. P. (2009). Electron backscatter diffraction in materials science. *Electron Backscatter Diffraction in Materials Science*. Springer, Boston, MA.

2. Zaefferer, S. (2011). A critical review of orientation microscopy in SEM and TEM. *Crystal Research and Technology*, *46*, 607–628.
3. Keller, R. R., & Geiss, R. H. (2012). Transmission EBSD from 10 nm domains in a scanning electron microscope. *Journal of Microscopy*, *245*, 245–251.
4. Fundenberger, J. J., Bouzy, E., Goran, D., Guyon, J., Yuan, H., & Morawiec, A. (2016). Orientation mapping by transmission-SEM with an on-axis detector. *Ultramicroscopy*, *161*, 17–22.
5. Sneddon, G. C., Trimby, P. W., & Cairney, J. M. (2016). Transmission Kikuchi diffraction in a scanning electron microscope: A review. *Materials Science and Engineering R: Reports*, *110*, 1–12.
6. Niessen, F., Burrows, A., & Fanta, A. B. da S. (2018). A systematic comparison of on-axis and off-axis transmission Kikuchi diffraction. *Ultramicroscopy*, *186*, 158–170.
7. Barsoum, M. L., Abbott, T. M., Jacobsen, S. D., Farha, O. K., & Dravid, V. P. (2026). Crystal Orientation and Defect Mapping in Electron-Beam-Sensitive Zeolites with Near-Axis Transmission Kikuchi Diffraction. *Nano Letters*. https://doi.org/10.1021/acs.nanolett.5c05800.
8. Ophus, C. (2019). Four-Dimensional Scanning Transmission Electron Microscopy (4D-STEM): From Scanning Nanodiffraction to Ptychography and Beyond. *Microscopy and Microanalysis*, *25*, 563–582.
9. Joy, D. C., Newbury, D. E., & Davidson, D. L. (1998). Electron channeling patterns in the scanning electron microscope. *Journal of Applied Physics*, *53*, R81.
10. Zaefferer, S. (2007). On the formation mechanisms, spatial resolution and intensity of backscatter Kikuchi patterns. *Ultramicroscopy*, *107*, 254–266.

11. Winkelmann, A., & Vos, M. (2011). Site-Specific Recoil Diffraction of Backscattered Electrons in Crystals. *Physical Review Letters*, *106*, 085503.
12. Winkelmann, A., & Vos, M. (2013). The role of localized recoil in the formation of Kikuchi patterns. *Ultramicroscopy*, *125*, 66–71.
13. Frank, F. C. (1972). A note on the reading of Kossel patterns. *Journal of the Less Common Metals*, *28*, 7–13.
14. Britton, T. B., Jiang, J., Guo, Y., Vilalta-Clemente, A., Wallis, D., Hansen, L. N., Winkelmann, A., & Wilkinson, A. J. (2016). Tutorial: Crystal orientations and EBSD — Or which way is up? *Materials Characterization*, *117*, 113–126.
15. Liu, Q. b., Cai, C. y., Zhou, G. w., & Wang, Y. g. (2016). Comparison of EBSD patterns simulated by two multislice methods. *Journal of Microscopy*, *264*, 71–78.
16. Winkelmann, A., Trager-Cowan, C., Sweeney, F., Day, A. P., & Parbrook, P. (2007). Many-beam dynamical simulation of electron backscatter diffraction patterns. *Ultramicroscopy*, *107*, 414–421.
17. Winkelmann, A., Nolze, G., Cios, G., Tokarski, T., Bała, P., Hourahine, B., & Trager-Cowan, C. (2021). Kikuchi pattern simulations of backscattered and transmitted electrons. *Journal of Microscopy*, *284*, 157–184.
18. Chen, Y. H., Park, S. U., Wei, D., Newstadt, G., Jackson, M. A., Simmons, J. P., De Graef, M., & Hero, A. O. (2015). A Dictionary Approach to Electron Backscatter Diffraction Indexing. *Microscopy and Microanalysis*, *21*, 739–752.
19. Foden, A., Collins, D. M., Wilkinson, A. J., & Britton, T. B. (2019). Indexing electron backscatter diffraction patterns with a refined template matching approach. *Ultramicroscopy*, *207*, 112845.

20. Ding, Z., Zhu, C., & De Graef, M. (2021). Determining crystallographic orientation via hybrid convolutional neural network. *Materials Characterization*, *178*, 111213.

21. Kaufmann, K., Zhu, C., Rosengarten, A. S., Maryanovsky, D., Harrington, T. J., Marin, E., & Vecchio, K. S. (2020). Crystal symmetry determination in electron diffraction using machine learning. *Science*, *367*, 564–568.

22. Vos, M., & Winkelmann, A. (2019). Effects of multiple elastic and inelastic scattering on energy-resolved contrast in Kikuchi diffraction. *New Journal of Physics*, *21*, 123018.

23. Zhang, T., Berners, L., Holzer, J., & Britton, T. B. (2025). Comparison of Kikuchi diffraction geometries in scanning electron microscope. *Materials Characterization*, *222*, 114853.

24. Brodu, E., Bouzy, E., & Fundenberger, J. J. (2017). Diffraction contrast dependence on sample thickness and incident energy in on-axis Transmission Kikuchi Diffraction in SEM. *Ultramicroscopy*, *181*, 123–133.

25. Pascal, E., Singh, S., Callahan, P. G., Hourahine, B., Trager-Cowan, C., & Graef, M. D. (2018). Energy-weighted dynamical scattering simulations of electron diffraction modalities in the scanning electron microscope. *Ultramicroscopy*, *187*, 98–106.

26. Krieger Lassen, N. C., Jensen, D. J., & Conradsen, K. (1992). Image Processing Procedures for Analysis of Electron Back Scattering Patterns. *Scanning Microscopy*, *6*.

27. Rauch, E. F., Portillo, J., Nicolopoulos, S., Bultreys, D., Rouvimov, S., & Moeck, P. (2010). Automated nanocrystal orientation and phase mapping in the transmission electron microscope on the basis of precession electron diffraction. *Zeitschrift fur Kristallographie*, *225*, 103–109.

28. Villert, S., Maurice, C., Wyon, C., & Fortunier, R. (2009). Accuracy assessment of elastic strain measurement by EBSD. *Journal of Microscopy*, *233*, 290–301.
29. Kacher, J., Landon, C., Adams, B. L., & Fullwood, D. (2009). Bragg's Law diffraction simulations for electron backscatter diffraction analysis. *Ultramicroscopy*, *109*, 1148–1156.
30. Dudarev, S. L., Peng, L.-M., & Whelan, M. J. (1993). Correlations in space and time and dynamical diffraction of high-energy electrons by crystals. *Physical Review B*, *48*, 13408–13429.
31. Dudarev, S. L., Rez, P., & Whelan, M. J. (1995). Theory of electron backscattering from crystals. *Physical Review B*, *51*, 3397–3412.
32. Shi, Q., Jiao, L., Loisnard, D., Dan, C., Chen, Z., Wang, H., & Roux, S. (2022). Improved EBSD indexation accuracy by considering energy distribution of diffraction patterns. *Materials Characterization*, *188*, 111909.
33. Schwarzer, R. A. (1997). Automated crystal lattice orientation mapping using a computer-controlled SEM. *Micron*, *28*, 249–265.
34. Britton, T. B., Maurice, C., Fortunier, R., Driver, J. H., Day, A. P., Meaden, G., Dingley, D. J., Mingard, K., & Wilkinson, A. J. (2010). Factors affecting the accuracy of high resolution electron backscatter diffraction when using simulated patterns. *Ultramicroscopy*, *110*, 1443–1453.

35. Zhang, T., & Britton, T. (2024). Multi-exposure diffraction pattern fusion applied to enable wider-angle transmission Kikuchi diffraction with direct electron detectors. *Ultramicroscopy*, *257*, 113902.

36. Qaiser, M. H., Berners, L., Scales, R. J., Zhang, T., Heller, M., Dluhos, J., Korte-Kerzel, S., & Britton, T. B. (2025, July 1). AstroECP: towards more practical Electron Channeling Contrast Imaging. arXiv.

37. Britton, T. B., Tong, V. S., Hickey, J., Foden, A., & Wilkinson, A. J. (2018). AstroEBSD: exploring new space in pattern indexing with methods launched from an astronomical approach. *Journal of Applied Crystallography*, *51*, 1525–1534.

38. Powell, C. (1999). NIST Electron Inelastic-Mean-Free-Path Database , NIST Standard Reference Database 71. National Institute of Standards and Technology. https://doi.org/10.18434/T48C78.

39. Winkelmann, A., Cios, G., Perzyński, K., Tokarski, T., Mehnert, K., Madej, Ł., & Bała, P. (2025). Simulation-based super-resolution EBSD for measurements of relative deformation gradient tensors. *Ultramicroscopy*, 114180.

40. Gauvin, R., Lifshin, E., Demers, H., Horny, P., & Campbell, H. (2006). Win X-ray: A New Monte Carlo Program that Computes X-ray Spectra Obtained with a Scanning Electron Microscope. *Microscopy and Microanalysis*, *12*, 49–64.

41. Bachmann, F., Hielscher, R., & Schaeben, H. (2010). Texture analysis with MTEX-Free and open source software toolbox. *Solid State Phenomena*, *160*, 63–68.

42. Singh, S., Ram, F., & De Graef, M. (2017). EMsoft: open source software for electron diffraction/image simulations. *Microscopy and Microanalysis*, *23*, 212–213.

43. Reimer, L., & Kohl, H. (2008). Transmission Electron Microscopy: Physics of Image Formation. Springer, New York, NY.

44. Vespucci, S., Naresh-Kumar, G., Trager-Cowan, C., Mingard, K. P., Maneuski, D., O'Shea, V., & Winkelmann, A. (2017). Diffractive triangulation of radiative point sources. *Applied Physics Letters*, *110*, 124103.

45. Nolze, G., Tokarski, T., & Rychłowski, Ł. (2023). Use of electron backscatter diffraction patterns to determine the crystal lattice. Part 1. Where is the Bragg angle? *Journal of Applied Crystallography*, *56*, 349–360.

46. Winkelmann, A. (2010). Principles of depth-resolved Kikuchi pattern simulation for electron backscatter diffraction. *Journal of Microscopy*, *239*, 32–45.

47. Brodu, E., & Bouzy, E. (2018). A New and Unexpected Spatial Relationship Between Interaction Volume and Diffraction Pattern in Electron Microscopy in Transmission. *Microscopy and Microanalysis*, *24*, 634–646.

Supplementary Information

# Dynamical Simulation of On-axis Transmission Kikuchi and Spot Diffraction Patterns, Based on Accurate Diffraction Geometry Calibration

Tianbi Zhang[1], Raynald Gauvin[2], Aimo Winkelmann[3,4], T. Ben Britton[1]

## Modular On-axis TKD Stage

The Si sensor of the direct electron detector is biased at 50 V (maximum allowed by manufacturer) to ensure full sensor depletion and high signal-to-noise. Only the minimum energy threshold of 3.01 keV (calibrated by the manufacturer) applied. There is no lens between the sample and the detector sensor.

Custom-made cables are used to connect the detector to a computer, and to connect the SmarAct stages to a control unit, which is also connected to the same computer. Capture of TKPs is performed with the Pixel Pro software (version 1.6.8)

## Dynamical Simulation

Table S1 lists selected parameters for the dynamical simulation (Table S1). Note that some parameters are unique to the simulation tool used in this study. Meaning of these parameters can be found in reference [1]. These factors remain unchanged across different types of simulations.

Table S1. Parameters used for dynamical simulation (based on the results shown in Figure 8 and 9).

| Physics-based parameters | | | |
|---|---|---|---|
| Inelastic mean free path (angstrom) | 155 | | |
| Debye-Waller Factor: crystal potential | 0.3 | Debye-Waller Factor: backscattering source | 0.3 |
| Other crystal structure information is read from the Crystallographic Information File (.cif) of the crystal lattice. | | | |
| Experimental parameters | | | |
| Beam energy (keV) | 30 | Pattern size (pixels) | 252x252 (CPE, IDI) |

| | | | 1024x1024 (Spots) |
|---|---|---|---|
| Pattern centre of Kikuchi pattern (corners of the gnomonic coordinates) [ $x_{min}$, $x_{max}$, $y_{min}$, $y_{max}$ ] | [ -0.33664, 0.50723, -0.40678, 0.43709] | | |
| Orientation (Euler angles, Bunge, °) | [114.6426, 99.9637, 166.2438] | | |
| Phenomenological parameters, pre-determined | | | |
| Maximum number of beams in the beam list | 6000 | Local beam list selection of relevant beams near each pixel | 6.0 |
| Minimum effective thickness (angstrom) | 0 | Maximum effective thickness (angstrom) | 1000 |
| Minimum lattice spacing (CPE, IDI) | 1.0 | Minimum structure factor relative to strongest reflection (CPE, IDI) | 0.35 |
| Minimum lattice spacing (spot pattern) | 0.05 | Minimum structure factor relative to strongest reflection (spot pattern) | 0.1 |
| Phenomenological scaling factor for the imaginary scattering potential | 0.1 | Phenomenological scaling factor for the real scattering potential | 1 |
| Strongest IDI intensity direction parameter $u_\varphi$ and $u_\theta$ (°) | $u_\varphi$=71.64° <br> $u_\theta$ =2.882° | | |
| | | | |
| Phenomenological parameters, fitted | | | |
| Spread of IDI intensity, $\alpha$ | | Weight factors for IDI and CPE patterns | |
| Critical energy loss threshold for | | | |

| discriminating coherent and incoherent scattering | | | |
|---|---|---|---|

Effect of downscaling the spot diffraction pattern is shown in Figure S1.

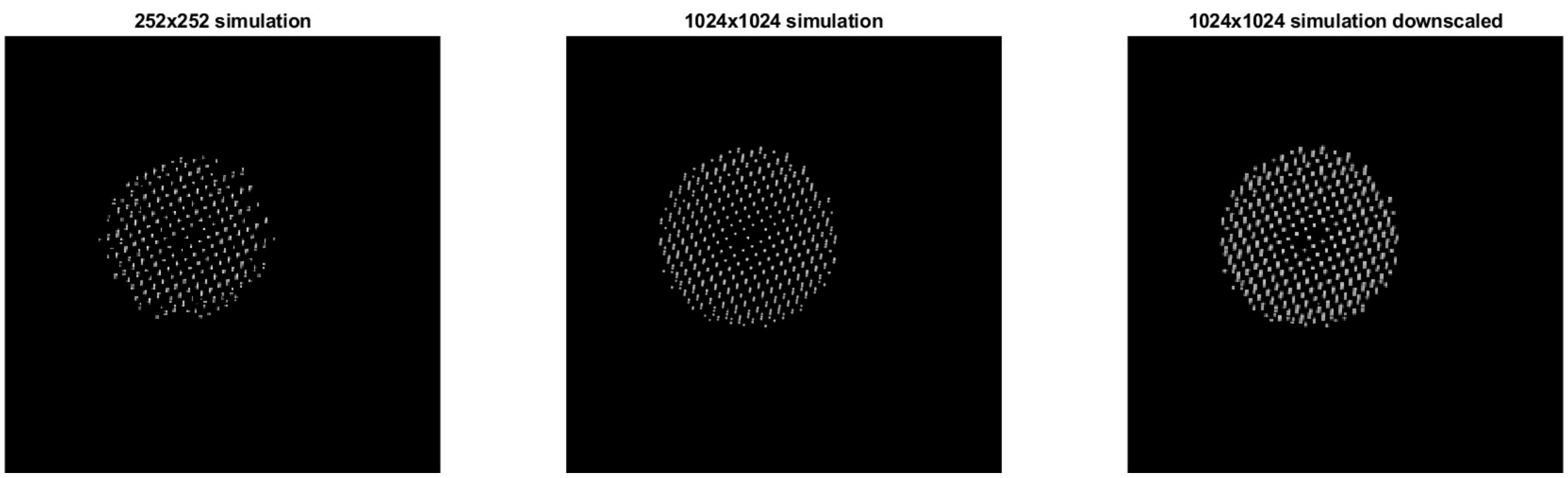

Figure S1. The effect of simulation size (252x252 and 1024x1024 pixels, and the latter down-scaled to 252x252) on the spot pattern.

**Benchmark of Geometry Simulation**

Figure S2 shows benchmark examples of the geometry simulation of spot patterns using zone axes patterns of γ -Fe at 20 keV.

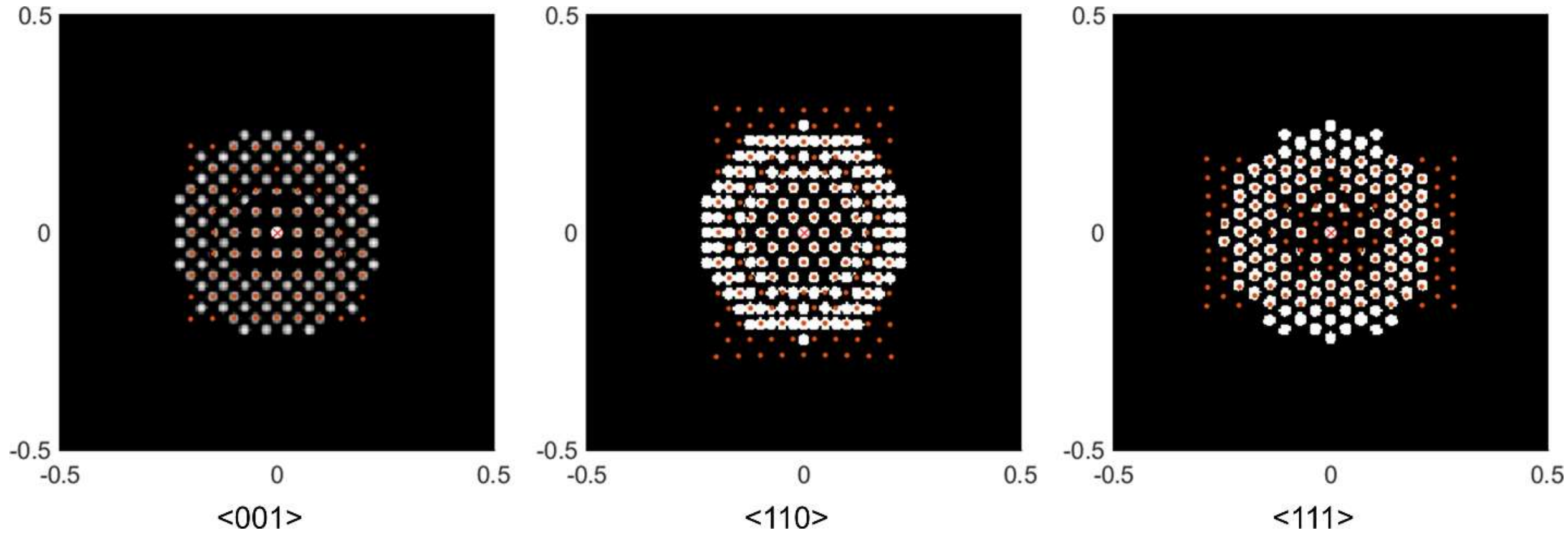

Figure S2. Dynamically simulated spot diffraction patterns of γ-Fe along selected zone axes at 20 keV, with kinematically simulated spots overlaid (orange). PC of the Kikuchi pattern is (0.5, 0.5, 1). No detector tilt.

## Effect of Tilt and in Geometric Simulation

The effect of incorporating the detector tilt is shown in Figure S3 and S4 using simulated and experimental patterns respectively. It is clear that detector tilt can strongly affect the appearance and accurate interpretation of on-axis TKD patterns, and the necessity of incorporating the shift correction.

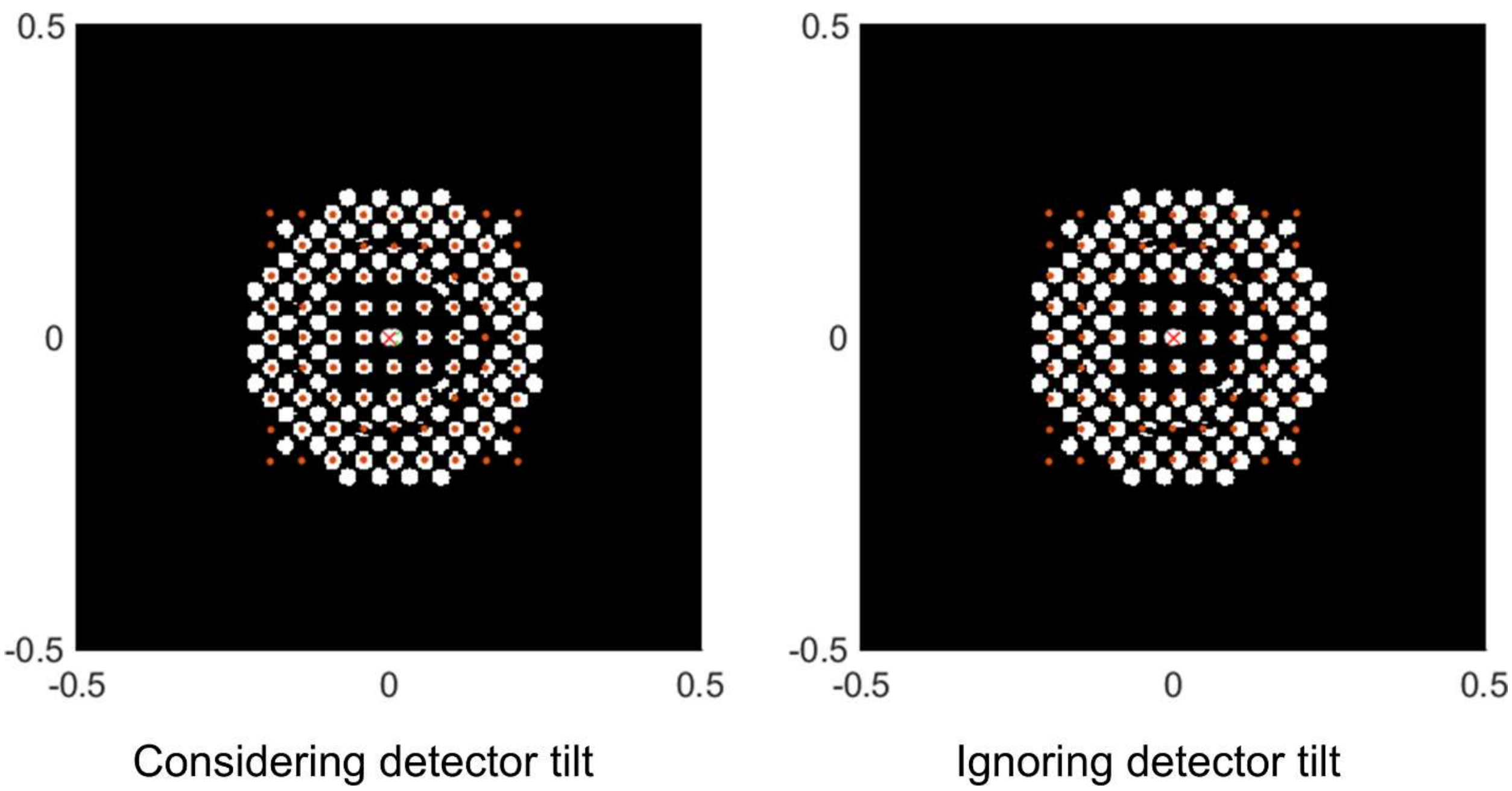

Figure S3. Dynamically simulated spot diffraction patterns of γ-Fe along zone axes at 20 keV, with kinematically simulated spots overlaid. The detector is tilted by 0.5° along Y - axis of the pattern. PC of the Kikuchi pattern (red) is (0.5, 0.5, 1) and the (000) spot is denoted by a green cross.

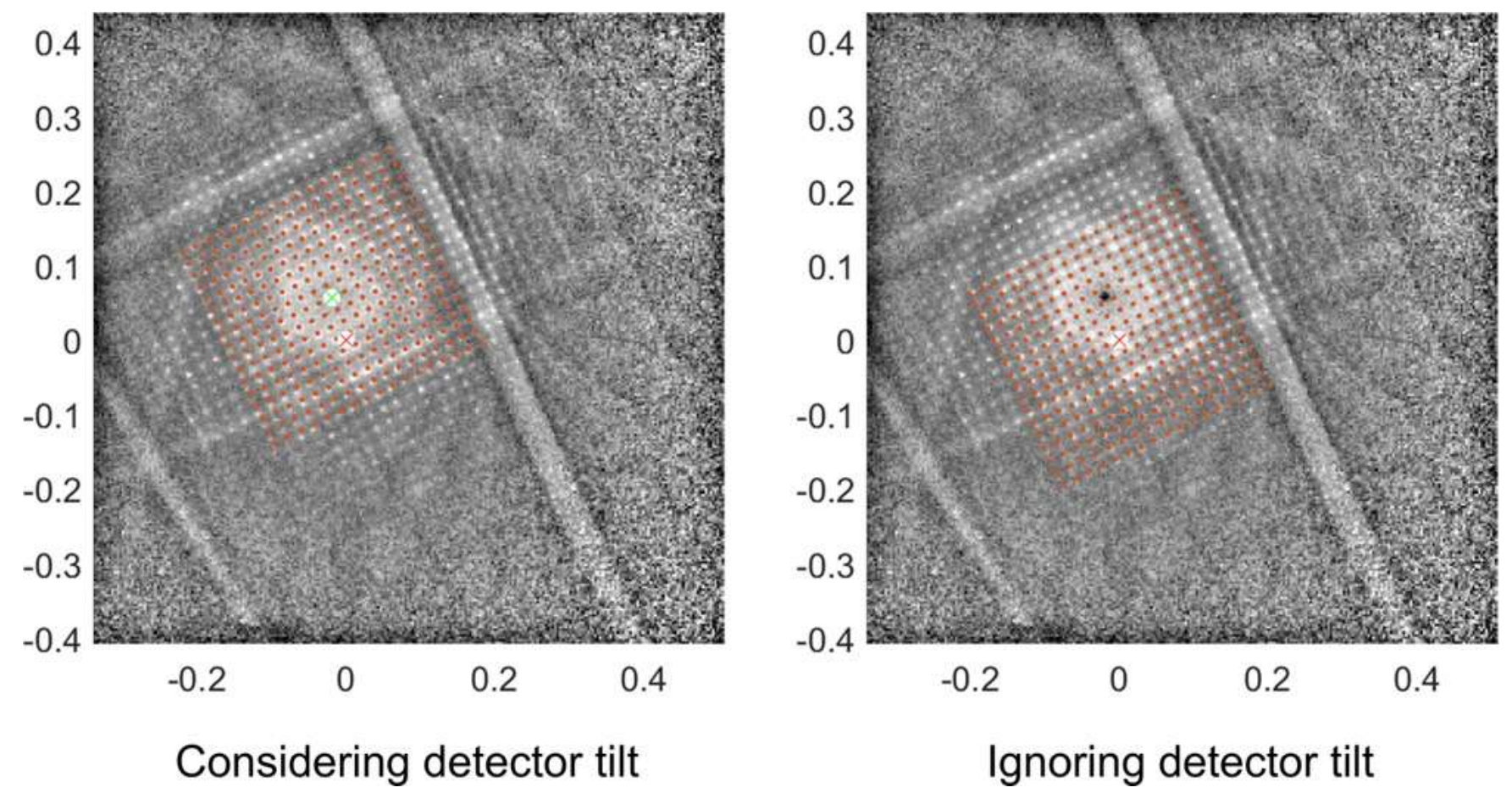

Figure S4. Experimental on-axis TKD patterns of $MoO_3$ at 30 keV with overlaid diffraction spots by considering and ignoring detector tilt.

## References

[1] A. Winkelmann, G. Nolze, G. Cios, T. Tokarski, P. Bała, B. Hourahine, C. Trager-Cowan, Kikuchi pattern simulations of backscattered and transmitted electrons, J. Microsc. 284 (2021) 157–184. https://doi.org/10.1111/jmi.13051.